\documentclass[reprint,
superscriptaddress,twocolumn, aps]{revtex4-2}
\usepackage[english]{babel}

\usepackage[letterpaper,top=2cm,bottom=2cm,left=3cm,right=3cm,marginparwidth=1.75cm]{geometry}

\usepackage{amsmath}
\usepackage{amsfonts}
\usepackage{graphicx}
\usepackage{siunitx}
\usepackage[colorlinks=true, allcolors=blue]{hyperref}
\usepackage{xr}

\begin{document}

\title{
Collective Behavior and Memory States in Flow Networks with Tunable Bistability}
\author{Lauren E.~Altman}
\email{laltman2@sas.upenn.edu}
\affiliation{Department of Physics \& Astronomy, University of Pennsylvania, Philadelphia, PA 19104, USA}
\author{Nadia~Aguilar}
\affiliation{Department of Physics, Princeton University, Princeton, NJ 08540, USA}
\author{Douglas~J.~Durian}
\affiliation{Department of Physics \& Astronomy, University of Pennsylvania, Philadelphia, PA 19104, USA}
\affiliation{Department of Mechanical Engineering and Applied Mechanics, University of Pennsylvania, Philadelphia, PA 19104, USA}
\author{Miguel~Ruiz-Garcia}
\affiliation{Departamento de Estructura de la Materia, Física Térmica y Electrónica, Universidad Complutense Madrid, 28040, Madrid, Spain.}
\affiliation{GISC - Grupo Interdisciplinar de Sistemas Complejos, Universidad Complutense Madrid, 28040, Madrid, Spain.}
\author{Eleni~Katifori}
\affiliation{Department of Physics \& Astronomy, University of Pennsylvania, Philadelphia, PA 19104, USA}
\affiliation{Center for Computational Biology,
Flatiron Institute,  New York, NY 10010, USA}

\begin{abstract}

Multistability-induced hysteresis has been widely studied in mechanical systems, but such behavior has proven more difficult to reproduce experimentally in flow networks. 
Natural flow networks like animal and plant vasculature can exhibit complex nonlinear behavior to facilitate fluid transport, so multistable flows may inform their functionality.
To probe such phenomena in an analogous model system, we utilize an electronic network of hysteretic bistable resistors designed to have tunable negative differential resistivity. 
We demonstrate our system's capability to generate complex global memory states in the form of voltage patterns, which is mediated by the tunable nonlinearity of each element's current-voltage characteristic.
We investigate avalanching behavior arising from effective interactions, and demonstrate how to encode explicit interactions of arbitrary form by taking advantage of the tunable circuitry design. 
\end{abstract}

\maketitle

\section{Introduction}
\label{sec:intro}

Return-point memory is a property of matter in which a system's response to external driving is a function of its history, such that the system retains some information of its previous state \cite{keim2019memory,paulsen2024memory}. 
The canonical example of a return-point memory system is the Preisach model for ferromagnets \cite{preisach1935magnetische}.
A ``hysteron" is a universal building block of return-point memory systems which can exist in multiple states, switch between states according to the external driving field, and exhibits a region of multistability. 
The addition of interaction terms to hysteron models opens pathways for higher-order effects such as self-loops \cite{baconnier2024proliferation} and the breakdown of return point memory \cite{lindeman2025generalizing, vanhecke2021profusion}.
Furthermore, some networks of physically-interacting bistable hysteretic elements demonstrate behaviors that are incompatible with the hysteron model entirely \cite{shohat2025geometric}.

There exist many mechanical implementations of such devices, ranging from crumpled paper \cite{bense2021complex, shohat2022memory} to balloons \cite{muhaxheri2024bifurcations,ilssar2020inflation,muller2004rubber,overvelde2015amplifying} to spring-bar systems \cite{paulsen2024mechanical,liu2024controlled} to buckling shells \cite{djellouli2024shell} to even knitted fabrics \cite{mahadevan2024knitting}.
Naturally occurring flow networks such as plant xylem and phloem \cite{park2021fluid} and rabbit cerebral arteries \cite{thorin1998high} exhibit bistable pressure-flow relationships resulting from fluid-structure interactions. 
Furthermore, model systems of microvasculature demonstrate that multistability may arise from nonlinear rheological effects such as viscosity contrast or hematocrit distribution \cite{alonzo2024spatio, geddes2010bistability}. 
Engineered \cite{martinez2024fluidic, rothemund2018soft, brandenbourger2020tunable, zhang2020self, gomez2017passive} flow networks that exhibit bistability are somewhat limited, and their hysteretic behavior is not typically exploited or reported.
Furthermore, it is often more difficult to produce this behavior in such fluidic systems because they require precise manufacturing as well as pressure control and measurement.

\begin{figure*}
    \centering
    \includegraphics[width=0.8\linewidth]{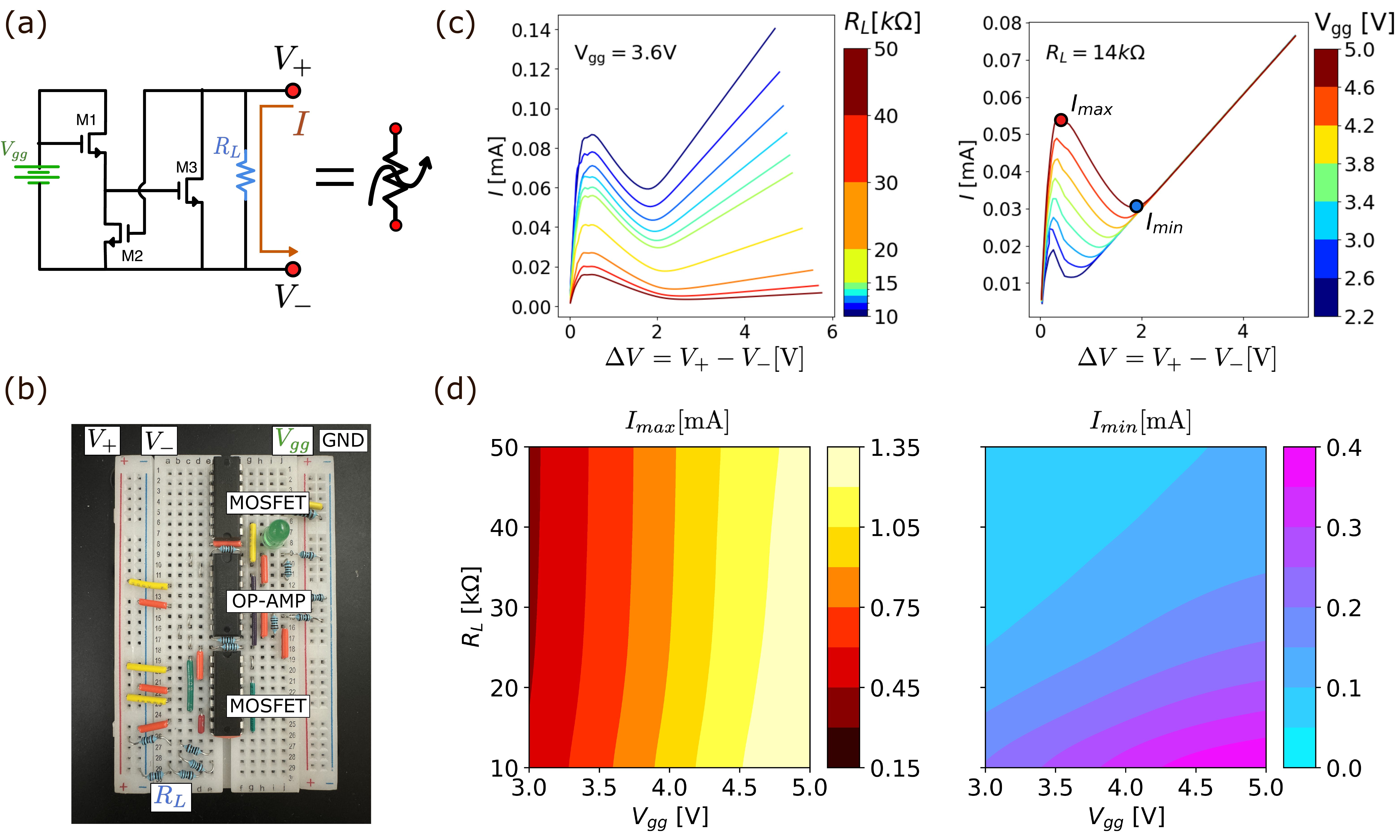}
    \caption{
    Construction of the NDR device.
    (a) Basic circuit diagram of NDR device highlighting the two tunable parameters, $V_{gg}$ (green) and $R_L$ (blue). In the rest of this work, the NDR device will be represented with a resistor overlaid with a wavy arrow.
    (b) Breadboard layout of the device.
    (c) IV characteristics of the NDR device for various values of $R_L$ (left), and $V_{gg}$ (right).
    (d) Heatmap showing the peak current $I_\textrm{max}$ and valley current $I_\textrm{min}$ as a function of $V_{gg}$ and $R_L$, based on experimental data.}
    \label{fig:tunableNDR}
\end{figure*}

While there are direct analogs between fluidic and mechanical networks in certain special cases, in general the mapping between such systems is not exact. 
This is owed to the fact that the conjugate variables in mechanical networks, namely forces and node positions, are vectors, while pressure and current in fluidic networks are scalars. 
A more appropriate mapping is to networks of linear resistors, 
\cite{kirchhoff1847ueber}, which have been successfully applied even to biological systems \cite{murray1926physiological}, although this mapping holds only in the low Reynolds number regime where inertial forces are negligible.
In this analogy, a resistor acts as a fluidic channel with its resistance corresponding to channel width.
Voltage drops across resistors represent pressure drops across channels, and the flow of charge (current) acts as a stand-in for fluid transport. Here we extend this analogy to a network of nonlinear resistors.
Nonlinear circuits typically do not have closed form solutions, but can produce richer behaviors than the linear model.
A device that exhibits a region of \textit{negative differential resistance} (NDR) is a special type of nonlinear element with a non-monotonic current-voltage relationship.
When combined with a linear resistor in series, the NDR region becomes unstable, so the two regions of positive differential resistance form the stable ``states" of this element. 

A variety of electronic devices exhibit hysteretic properties, including memristors \cite{strukov2008missing} and Gunn diodes \cite{shiri2018gunn}, which are semiconductor devices that operate via the Gunn effect.
Each of these devices exhibit nonlinear current-voltage flow characteristics, but  
tuning these systems for controlled effects poses engineering challenges. 
Metal-Oxide Field Effect Transistor (MOSFET)-based NDR devices \cite{chen2009negative, ulansky2019electronic} are in contrast low-cost, easy to construct, and widely tunable.
The memory capabilities of such devices have been previously reported \cite{lee2020analysis}, but the emergent memory effects arising from networks of these devices has yet to be investigated.

In this work, we construct a FET-based NDR device and show how the device's current-voltage relationship, hereafter referred to as its ``IV curve," can be tuned by modifying two of its internal properties independently.
This tunability is essential for utilizing the full memory capabilities of networks of these devices, as the relative properties of each edge determine the accessible states the network can achieve through emergent hysteresis.
We demonstrate how a network of these devices can encode memories of the history of driving by a global source voltage.
We also observe avalanching behavior and demonstrate how this can be explained and predicted from geometric interactions \cite{shohat2022memory}.
Finally, we present a method for encoding explicit interactions between devices by controlling the device's tunable IV characteristic according to the network state.
The versatility of our device construction for encoding interactions of arbitrary form makes it an ideal platform for probing hysteretic network behaviors, and hints at opportunities for engineered networks with sustained oscillations \cite{ruiz2021emergent} and multiperiodic orbits \cite{keim2021multiperiodic}.

\section{Results}
\subsection{Tunable Nonlinearites}
\label{sec:tunable}

An edge in a circuit network is connected to two nodes, $V_{+}$ and $V_{-}$, and experiences a current $I$ flowing through it. 
Our NDR device, shown in a simplified schematic in Fig.~\ref{fig:tunableNDR}(a), exhibits a non-monotonic relationship between the current and the voltage drop across the edge, $\Delta V = V_+ - V_-$. 
The non-monotonic behavior arises as a result of the interaction between the three MOSFETs \cite{sarkar2016dc}.
Initially, transistor M3 is in triode mode while transistors M1 and M2 are in cutoff mode, so the entire device acts as a linear resistor.
At $\Delta V > V_{th}$, with $V_{th}$ as the threshold voltage, transistor M2 will turn on and be in saturation mode, while M3 is still in triode mode, resulting in a current voltage relationship $I(\Delta V) \propto \Delta V - \Delta V^2$.
At $\Delta V > \frac{1}{2}(V_{gg} - V_{th})$, M3 enters saturation mode, resulting in a relationship of the form $I(\Delta V) \propto -\Delta V^2$.
The linear resistor in parallel, $R_L$, provides the second positive differential resistance region and therefore the bistability of the device.
All current between nodes $V_+$ and $V_-$ flows through either M3 or the linear resistor $R_L$, so for a given constant $V_{gg}$, this is a two-terminal device. 

The breadboard layout of the device, depicted in Fig.~\ref{fig:tunableNDR}(b), includes two MOSFET chips (ALD1106PBL) and one Operational-Amplifier (Op-Amp) chip (TLV274CDR), which is used for supplying the gate voltage $V_{gg}$. 
Included on each board is an LED whose brightness is proportional to $\Delta V$.
The full circuit diagram is depicted in Materials and Methods, Fig.~\ref{fig:circuit}.

\begin{figure*}
    \centering
    \includegraphics[width=\linewidth]{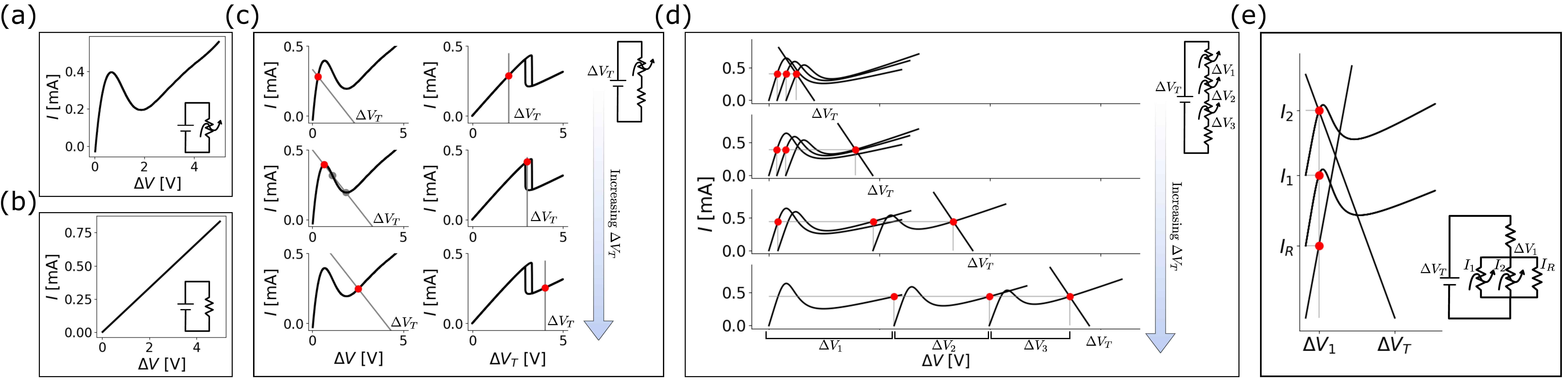}
    \caption{
    Geometric solutions of small network interactions between linear and nonlinear elements.
    (a) Native IV curve for an NDR device.
    (b) Native IV curve for a linear resistor, $R_0 = \SI{6}{k \Omega}$.
    (c) Solutions for current in a network with one linear resistor and one nonlinear resistor in series, for various values of $\Delta V_T$. The left column shows the geometric solution of Eq.~\ref{eq:geometric} and the right column the $I(\Delta V_T)$ for the NDR plus linear resistor circuit.
    The total $I(\Delta V_T)$ curve shows a hysteresis loop.
    (d) Solutions for current in a network with one linear resistor and three nonlinear resistors in series, for various values of $\Delta V_T$.
    The ordering that each nonlinear element switches state is dependent on each curve's peak current, $I_\textrm{max}$.
    (e) Geometric construction of a branching network. Elements in parallel are stacked vertically, and elements in series are stacked horizonally.}
    \label{fig:geometric}
\end{figure*}

A key feature of this device is its strong tunability. 
Two internal properties, the gate voltage of M1 $V_{gg}$, and the linear resistance $R_L$, can be varied independently. 
Each tuning knob has an effect on the IV curve of the device, as shown in Fig.~\ref{fig:tunableNDR}(c).
As $V_{gg}$ increases with $R_L$ held fixed, the height of the peak of the curve increases, and as $R_L$ increases with $V_{gg}$ held fixed, the slope of the second positive differential resistance region decreases.

The switching behavior of the device is predominantly dictated by the extrema of the current along the device's IV curve: the current at the peak, $I_\textrm{max}$, and the current at the valley, $I_\textrm{min}$.
Fig.~\ref{fig:tunableNDR}(d) summarizes how our two tuning knobs influence these values in a heatmap. 
The current at the peak $I_\textrm{max}$ is primarily dictated by $V_{gg}$, while the current at the valley $I_\textrm{min}$ is strongly influenced by both $V_{gg}$ and $R_L$. 

\subsection{Small network interactions}
\label{sec:interaction}

Even small networks consisting of NDR devices and linear elements are capable of emergent hysteretic behavior. 
Figs.~\ref{fig:geometric}(a) and (b) depict typical IV characteristics of an NDR device and a linear resistor, respectively.
The NDR device obeys a nonlinear IV curve, $I = \Gamma(\Delta V_\textrm{NDR})$,
and the linear resistor obeys the usual Ohm's law, $\Delta V_R=IR$.
When these elements are placed in series with each other at a DC source voltage $\Delta V_T$, as shown in Fig.~\ref{fig:geometric}(c, inset), one may use Kirchoff's laws to analytically describe the behavior of the network.
The total voltage across the system must be the sum of the voltage drops across each element,
\begin{equation}
\label{eq:voltage}
    \Delta V_T = \Delta V_\textrm{NDR} + \Delta V_{R},
\end{equation}
and the current through each element must be equivalent,
\begin{equation}
\label{eq:current}
   I_\textrm{NDR} = I_R = I.
\end{equation}
Eqs.~\eqref{eq:voltage} and~\eqref{eq:current}, together with the IV characteristics of each element, can be solved to obtain a governing equation for the system,
\begin{equation}
\label{eq:geometric}
    I = \frac{\Delta V_T - \Delta V_\textrm{NDR}}{R} = \Gamma(\Delta V_\textrm{NDR}).
\end{equation}

Fig.~\ref{fig:geometric}(c) depicts this governing equation geometrically.
In the left plots, the black curve represents the nonlinear IV characteristic, $\Gamma(\Delta V_\textrm{NDR})$, while the gray line represents the left-hand side of Eq.~\eqref{eq:geometric}.
The right plots depict the total $I(V)$ curve for the combined network of linear and nonlinear element, which includes a hysteresis loop. 
In correspondence with the left plots, the gray line represents the instantaneous total voltage, $\Delta V_T$, and is vertical in this case.
In both cases, the red point denoting the intersection between gray and black lines is the solution to Eq.~\eqref{eq:geometric} and therefore governs the physical behavior of the network.
The combined network of linear and nonlinear resistor together form the object known as a hysteron.

It should be noted that this network configuration may or may not exhibit multistabilities depending on the interaction between the properties of each element with each other. 
When $R$ is sufficiently small,
the gray lines in Fig.~\ref{fig:geometric}(c, left) are nearly vertical, so there may never be multiple solutions to Eq.~\eqref{eq:geometric}.
In this case, the system will sit stably in the negative differential resistance region of $\Gamma (\Delta V_\textrm{NDR})$ \cite{martinez2024fluidic}. 
This highlights a key deviation between our basic network element and a hysteron.
The critical resistance value $R_C$ for achieving multistable behavior can be computed by plugging the local minima and maxima of $\Gamma (\Delta V)$ into Eq.~\ref{eq:geometric}, resulting in
\begin{equation}
    R_C = \frac{V_{min} - V_{max}}{I_{max} - I_{min}},
\end{equation}
where $V_{max}, I_{max}$ are the voltage and current at the peak of the nonlinear IV curve, and $V_{min}, I_{min}$ are the voltage and current at the valley.

We now consider the case $R > R_C$.
At low $\Delta V_T$, there is one solution to the equation, dictating that the NDR device sit in its low-voltage stable state. 
As $\Delta V_T$ increases, the system may enter into a multistable region, where there are three solutions to Eq.~\eqref{eq:geometric}.
Two of these solutions are stable, associated with either the low-voltage or high-voltage states, and the solution which sits in the region of negative differential resistance is unstable \cite{martinez2024fluidic}. 
At high values of $\Delta V_T$, there is again only one solution to Eq.~\eqref{eq:geometric}, dictating that the NDR device sit in its high-voltage stable state. 
The nonlinear device's transition from low- to high-voltage state when $\Delta V_T$ is increasing occurs at the local current maximum in the multistable region, whereas the transition from high- to low-voltage state when $\Delta V_T$ is decreasing occurs at the local current minimum.
The animation depicted in Supplemental Video~1 illustrates this relationship in an experimental example.
While we model this as a near-discontinuous transition between states, in reality, small parasitic capacitances in our components result in transition timescales that are faster than our measurement capabilities. 
For more information about capacitance, see Supplementary Information, Section~\ref{sec:capacitance}.

\begin{figure*}
    \centering
    \includegraphics[width=\linewidth]{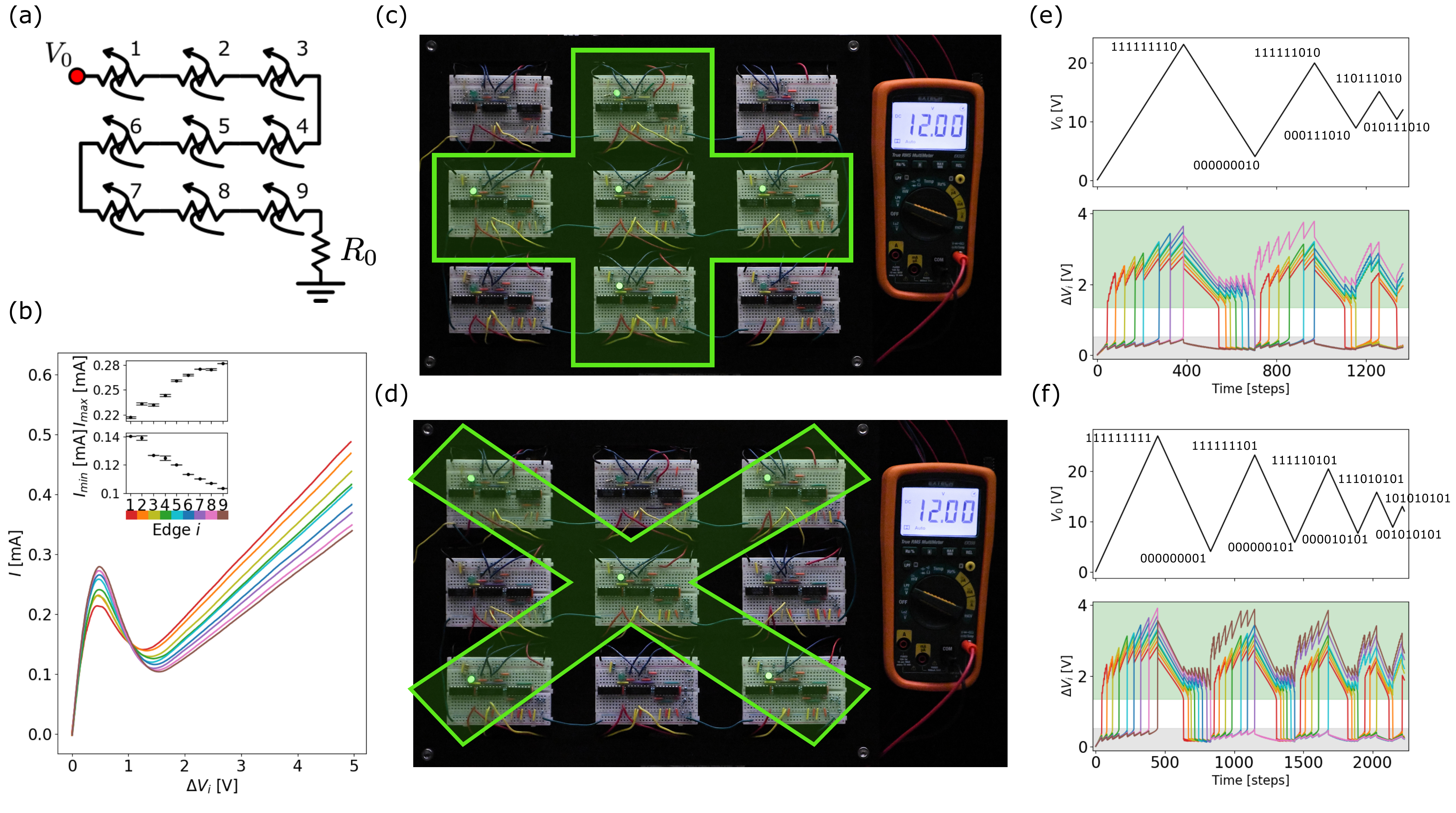}
    \caption{
    Demonstration of emergent memory effects in a network of 9 NDR elements in series.
    (a) Diagram showing network connectivity. The network is held at a total voltage $V_0$, and current is measured with a small resistor $R_0 = \SI{100}{\Omega}$.
    (b) The IV curves of each of the nonlinear elements have been tuned with respect to each other to control the ordering of state switching. Inset: nominal values of $I_\textrm{max}$ and $I_\textrm{min}$ for each edge.
    (c-d) Photographs of the network in the states ``010111010", which visually represents a ``$+$", and ``101010101", which visually represents an ``$\times$", respectively. The multimeter measures a global drop of $V_0 = \SI{12}{V}$ in both cases. 
    (e-f) Histories of total source voltage ramping $V_0(t)$ (top), as well as the voltage drops of each nonlinear edge $\Delta V_i$ (bottom), to achieve the states ``$+$" and ``$\times$", respectively. Extrema of $V_0$ are labeled with the binary string achieved at the given time step.
    }
    \label{fig:9NDR}
\end{figure*}

Similar geometric constructions may be applied for other minimal networks. 
Fig.~\ref{fig:geometric}(d) depicts three nonlinear devices and one linear resistor in series.
Each nonlinear element obeys an IV characteristic of the form $I = \Gamma_i(\Delta V_i)$, though each $\Gamma_i$ need not be identical.
The constraint that the current through each element must be equivalent is represented by a horizontal line through each IV characteristic.
The sum of voltage drops across all elements must sum to $\Delta V_T$, so the IV characteristic of each successive element $\Gamma_{i+1}(\Delta V_{i+1})$ is shifted by $\Delta V_i$, the voltage value that satisfies the current constraint.
The order in which elements transition between states as $\Delta V_T$ increases or decreases is set by the height of the maxima and minima of their IV characteristics.

Fig.~\ref{fig:geometric}(e) involves a circuit with elements both in series and in parallel. 
Elements in parallel are stacked vertically along the voltage-current plot, owing to the constraint of a constant voltage drop across them.
These parallel elements may then be stacked horizontally with the linear element in series.
The simple rules of this construction can be extended to networks of arbitrary topology and size.

It should be noted that whereas a single NDR and linear resistor in series could exhibit bistability and hysteresis, behaviors consistent with an isolated hysteron, a number of NDRs in series with a resistor cannot be thought of as a collection of interconnected hysterons, as a single NDR is not a hysteron.  

\subsection{Emergent network memories}
\label{sec:emergent}

The interaction of multiple NDR elements produces an emergent system-scale memory that is encoded in the binary voltage state value of each element \cite{ferrari2022preisach,terzi2020state}. 
Elements at low voltage are represented as the ``0" state, and the LED on that element's breadboard will be off, while elements at high voltage are represented as the ``1" state, and the LED will be on.
The point at which a transition between states occurs is strongly dependent on the properties of that element's IV characteristics, as well as the properties of all other elements in the network.
A series circuit of \num{9} elements  is schematically depicted in Fig.~\ref{fig:9NDR}(a).
The network will be driven only by the global source voltage $V_0$, and the ordering of transition states can be determined using the values of $I_\textrm{max}$ and $I_\textrm{min}$ of each element \cite{muhaxheri2024bifurcations}.

Note that the linear resistor $R_0$ is present for current measurement purposes and does not play a role in the network's behavior in this case. Whereas in the single resistor case a resistor in series is necessary to achieve hysteretic behavior, two or more NDRs in series can collectively exhibit multistability and as a consequence of memory effects. 

To construct a network in which every possible binary state is accessible by global driving, elements should be tuned relative to each other such that each successive element has a higher $I_\textrm{max}$ and lower $I_\textrm{min}$ than the previous \cite{teunisse2024transition},
\begin{equation}
\label{eq:current_order}
    I_\textrm{max}^{i+1} > I_\textrm{max}^i, \,
    I_\textrm{min}^{i+1} < I_\textrm{min}^i.
\end{equation}
The transition states are described explicitly for the three-edge network depicted in Fig.~\ref{fig:geometric}(d) in Supplementary Information, Section~\ref{sec:preisach}.
The nine IV curves depicted in Fig.~\ref{fig:9NDR}(b) obey Eq.~\ref{eq:current_order}, and the values of $I_\textrm{max}$ and $I_\textrm{min}$ for each edge are plotted in Fig.~\ref{fig:9NDR}(b, inset).
In this case, the ordering corresponds to the physical position in the series circuit, but this ordering is independent of physical position, and depends only on the relative IV characteristics of each edge.
With this ordering, when the total voltage $V_0$ across the circuit increases, each $i^{th}$ edge in the ``off" state will transition into its ``on" state before the ${i+1}^{th}$, and when the total voltage decreases, each $i^{th}$ edge in the ``on" state will transition into its ``off" state before the ${i+1}^{th}$.
This ordering is demonstrated in Supplemental Video~2.
We may then define an iterative algorithm for obtaining any binary string of on/off states. 

As a demonstration of this concept, Figs.~\ref{fig:9NDR}(c) and (d) depict the same network in two distinct binary states: \textit{010111010}, which visually looks like a ``$+$'', and \textit{101010101}, which visually looks like an ``$\times$''.
Both networks are held at a constant source voltage of $V_0 = \SI{12}{V}$
and the properties of each individual edge are identical between the two pictures.
The difference in their states can be attributed solely to the history of $V_0(t)$ for each state, depicted in Figs.~\ref{fig:9NDR}(e-f).
The annotations at the extrema of the $V_0$ history mark the binary state obtained at that point in the sequence, and the bottom plots corroborate this with exact voltage drop values for each edge, $\Delta V_i$. 
Voltages in the green regions are considered to be in the ``on" state, while those in the gray regions are considered ``off". 
The white regions are unstable and not experimentally accessible in this configuration.
In both cases, the final branch of $V_0$ ramping is present to achieve the same exact voltage of $V_0 = \SI{12}{V}$, but does not change the binary states of the network.
Supplemental Video~2 depicts the full ramping procedure for both states.

\begin{figure*}
    \centering
    \includegraphics[width=\linewidth]{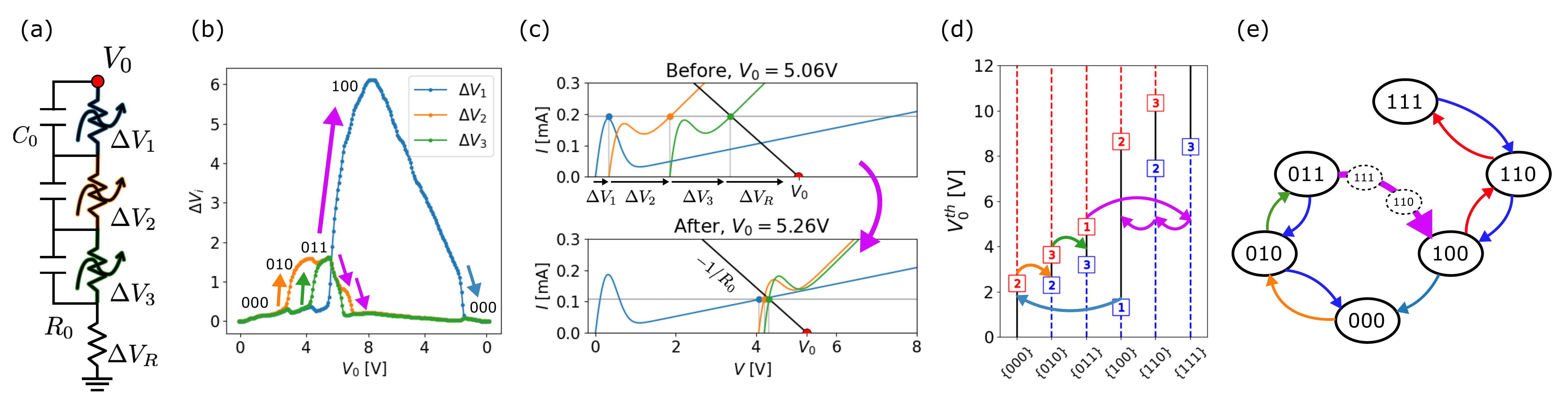}
    \caption{
    Avalanches arise from effective network interactions.
    (a) Schematic of the network. Three NDR edges are arranged in series with one linear resistor, and the network is controlled via the global source voltage, $V_0$.
    (b) Voltage drops of each edge as $V_0$ is ramped up and then back down. Arrows indicate where edge transitions occur, and states are labeled by their binary sequence. The three purple arrows denote the avalanche.
    (c) Geometric construction of the network just before (top) and just after (bottom) the avalanche. 
    (d) Transition thresholds and stability of each binary state, obtained from SPICE simulation. Colored arrows indicate the transitions corresponding to the arrows in (b). The avalanching transition passes through two unstable states, \textit{111} and \textit{110}.
    (e) Full transition graph for the network with tunings shown.
    }
    \label{fig:avalanche}
\end{figure*}

Because our elements are in series and under voltage control, there exist effective interactions between elements that result in transition thresholds for each edge that depend on the state of the network.
These can be observed explicitly in Fig.~\ref{fig:9NDR}(e-f) in the sharp changes in edge voltage drops $\Delta V_i$ when another edge makes a transition.
For a number of reasons, including the relatively large number of elements \cite{teunisse2024transition,shohat2025geometric} and the relatively narrow window of parameter tunings across all edges, these interactions are small and do not affect the transition graph of the network.
Supplemental Information, Section~\ref{sec:9ndr_interactions} includes explicit measurements of transition thresholds, as well as a geometric argument for why these interactions are not strong enough to affect the transition graph.
The following section involves a more detailed treatment of these effective interactions.

\subsection{Effective Interactions}
\label{sec:avalanche}

Because our bistable units are constrained by physics, namely, Kirchoff's laws, effective interactions between edges arise \cite{shohat2025geometric}.
In this case, our control parameter is the total voltage across a series circuit, $V_0$, which must equal the sum of voltage drops across all edges.
These physically-mediated interactions represent a divergence from the hysteron model, and can lead to behavior that violates its assumptions.

Such effective interactions allow for avalanches in which multiple edges transition states as a result of cascading instabilities.
Fig.~\ref{fig:avalanche}(a) depicts a network of three nonlinear edges and one linear resistor with $R_0 = \SI{8720}{\Omega}$ in series under voltage control. 
Capacitors with $C_0 = \SI{22}{\mu F}$ have been added in parallel to each nonlinear edge to slow the equilibration dynamics for measurement purposes.
See Supplements, Sec.~\ref{sec:capacitance} for more information about the role of capacitance.

Fig.~\ref{fig:avalanche}(b) shows the progression of states and voltage drops for each edge over the course of the experiment.
This network undergoes an avalanche of the type ``up-down-down,"
in which the first transition triggers instabilities in the other two edges.
When $V_0$ is ramped up from \SI{0}{V} to \SI{8}{V}, edges 2 (orange arrow) and 3 (green arrow) first undergo ``up" transitions, placing the network in the state \textit{011}.
Shortly thereafter, edge 1 transitions upward, but requires such a large voltage drop to do so that it forces edges 2 and 3 to drop back down (purple arrows), resulting in the state \textit{100}.
With the slowed-down relaxation timescales, we can see clearly the ordering of switching events \cite{jin2025dynamic}, indicating that these effects are dynamical in nature.

To corroborate our experiment, we construct a SPICE simulation of the network using IV characteristics that are parameterized from experimental data according to the form 
\begin{equation}
    I(\Delta V) = A \Delta V\,  e^{-\beta (\Delta V-V_0)^2} + k_0 \Delta V,
\end{equation}
where $A$, $\beta$, $V_0$, and $k_0$ are fitting parameters.
Fig.~\ref{fig:avalanche}(c) shows the geometric construction for the simulated network's state just before (top) and just after (bottom) the avalanche event. 
Graphically solving for the network constraints, we see that just after edge 1 transitions up, the only stable solution for edges 2 and 3 are in their lower branch.
The IV curves of each edge paint a picture of the necessary tunings to obtain avalanching behavior. 
Edge 1's IV curve has a significantly lower $I_{min}$ and larger $R_L$ than edges 2 and 3, indicating that it requires a much greater voltage drop to sit in its upper branch.

Fig.~\ref{fig:avalanche}(d) depicts the transition thresholds of the global switching field, $V_0^{th}$ arranged by binary state, obtained from SPICE simulation.
Solid bars indicate a region of stability, while dashed lines indicate that the state would be unstable for that given value of $V_0$. 
A red (blue) boxed number for a given network state indicates that the edge with the corresponding number will be the next to transition up (down), and boxes are plotted according to the transition threshold for that edge, $V_0^{th}$.
The boxed transitions separate the stable and unstable regions.
Only the experimentally accessible transition thresholds have been plotted, ie. the lowest ``up" threshold (red) and the highest ``down" threshold (blue) for each state.
Two states, \textit{101} and \textit{001}, are also not shown, as they are not accessible via transitions from the saturating states.

The colored arrows between bars correspond to the transitions shown in Fig.~\ref{fig:avalanche}(b). 
The purple arrow from \textit{011} shows that the resulting state at \textit{111} is unstable, with edge 3 being most inclined to transition down, so the network moves to \textit{110}. 
However, this state is also unstable, so edge 2 switches down. 
Finally, the network settles in the \textit{100} state, which is stable at the given $V_0$. 
The cascading behavior and switching order can be seen explicitly in the slowed-down dynamics of Fig.~\ref{fig:avalanche}(b), although the voltages at which switches occur are different than those obtained from the simulation that operates in the quasistatic limit.

Fig.~\ref{fig:avalanche}(e) summarizes the transition sequence of the network that can be obtained from the information in Fig.~\ref{fig:avalanche}(d).
The passage from \textit{011} to \textit{100} through the two intermediate unstable states of \textit{111} and \textit{100} are depicted as bubbles with dashed borders.

\subsection{Encoded Interactions}
\label{sec:encoded}

\begin{figure*}
    \centering
    \includegraphics[width=\linewidth]{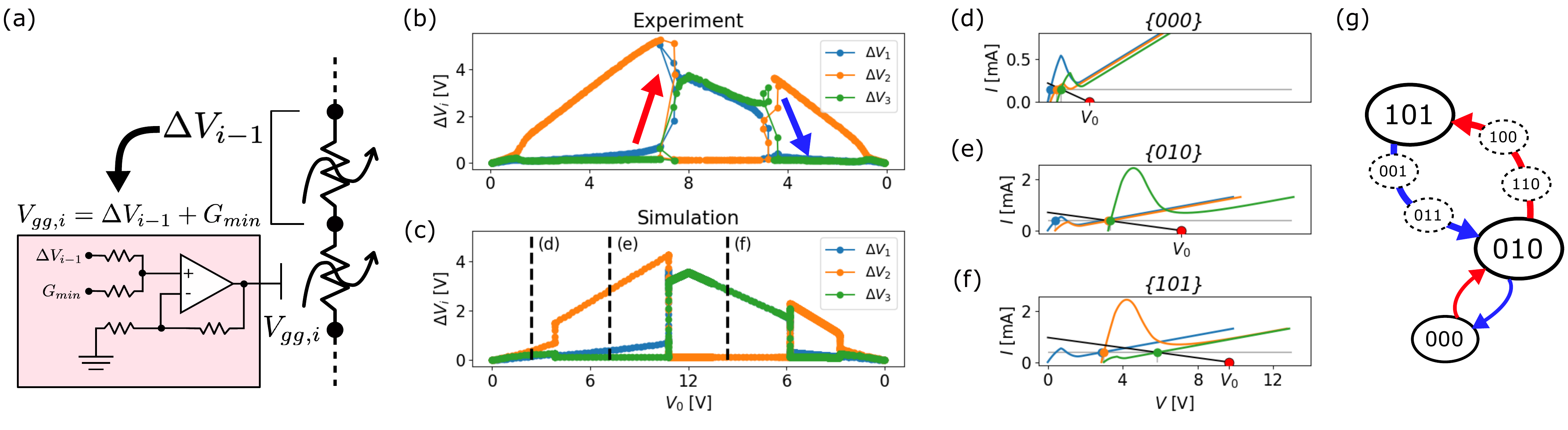}
    \caption{
    Encoded anti-cooperative interactions result in antiferromagnetic behavior.
    (a) The gate voltage of each edge, $V_{gg,i}$, is a function of the voltage drop of its neighbor, $\Delta V_{i-1}$.
    The boxed area shows the voltage adder circuitry which results in $V_{gg,i} = \Delta V_{i-1} + G_\textrm{min}$.
    Experimental data (b) and analogous SPICE simulation (c) demonstrate the network's antiferromagnetic response.
    (b) Voltage drops of each edge as a function of the global drive, $V_0$, as it is ramped up and down. Arrows indicate the avalanching transitions from \textit{010} to \textit{101} (red) and from \textit{101} to \textit{010} (blue).
    (c) Dashed lines indicate snapshots of the simulated network at different binary states in (d-f).
    The geometric constructions of each snapshot in (c) show how the IV curves of the edges evolve from the \textit{000} (d) to \textit{010} (e) to \textit{101}  (f) states.
    (g) The transition graph for this network summarizes the antiferromagnetic avalanches.}
    \label{fig:encoded}
\end{figure*}

Our experimental platform lends further advantages as compared with other hysteretic systems because we can leverage the device's tunability to encode explicit interactions between edges. 
Unlike the interactions in Sec.\ref{sec:avalanche}, which arise from the network topology and the method of driving, these encoded interactions can be explicitly designed to alter the IV characteristics of each edge as a function of the voltage drop of the others. 
In this way, we can incorporate localized spatial dependence into the network's behavior, a feature which has not been present in any of the previous demonstrations in this work.

We achieve this by tying the gate voltage of each edge to some function of the voltage drops of its neighbors. 
Fig.~\ref{fig:encoded}(a) shows schematically one example of how this can be achieved.
In a series network, the gate voltage of edge $i$, $V_{gg, i}$, linearly depends on the voltage drop of the edge that comes before it in the series, $\Delta V_{i-1}$.
\begin{equation}
\label{eq:AFM}
    V_{gg,i} = \Delta V_{i-1} +  G_\textrm{min}
\end{equation}
The constant offset, $G_\textrm{min} = \SI{2}{V}$, ensures that the IV curve of each edge remains bistable regardless of the state of the network.
The function in Eq.~\ref{eq:AFM} is implemented using the voltage summation circuit in the highlighted box, though this circuitry can be replaced to achieve any number of functional relationships between voltage drops and gate voltages.

Figs.~\ref{fig:encoded}(b-c) show experimental results for a network with encoded interactions. 
The network includes three nonlinear edges and a linear resistor, $R_0 =\SI{994}{\Omega}$, in series under voltage control. 
Edges \num{2} and \num{3} include the gate voltage relationship shown in (a), as can be seen in the plots in Fig.~\ref{fig:encoded}(c).
Edge \num{1} holds a constant gate voltage, $V_{gg,1} = \SI{3.55}{V}$.
Capacitors with $C_0 = \SI{22}{\mu F}$ have been placed in parallel with the nonlinear edges to slow the dynamics for measurement.

The relationship between $\Delta V_{i-1}$ and $V_{gg,i}$ results in anti-cooperative interactions between adjacent edges, and therefore a globally antiferromagnetic behavior of the network.
This can be seen experimentally in Fig.~\ref{fig:encoded}(b), which shows the voltage drops of each edge as the global drive, $V_0$, is ramped up and then back down.
From the state \textit{010}, edge \num{1}'s upward transition triggers an avalanche.
Edge \num{1}'s rapidly increasing voltage drop leads to a sharp increase in the height of the peak of edge \num{2}, causing it to drop back down. 
This event then cascades to edge \num{3}, which now has a much smaller peak and can transition upward. 
The network settles finally in the \textit{101} state, and follows the same pathway when $V_0$ is ramped down.
We note here that although both instances involve avalanches of the form ``up-down-down," the demonstration here differs from that depicted in Fig.~\ref{fig:avalanche}(b), owing to the varying IV curves.
This can be observed during the ramp-down procedure, which results in no avalanche in the case of effective interactions, but another ``down-down-up" transition in the case of explicit interactions.

Fig.~\ref{fig:encoded}(c) shows SPICE simulation data for a network with the same structure as the experiment, with $V_{gg,1} = \SI{2.50}{V}$ and $G_\textrm{min} = \SI{1.75}{V}$. 
Three vertical dashed lines indicate time slices of the network at different points in the progression, each of which are plotted as geometric constructions in (d-f).
In the \textit{010} state, edge \num{2} has a low peak due to edge \num{1} being at low voltage. 
Since edge \num{2} is in state \textit{1}, its voltage drop is high, and the peak of edge \num{3} is also high, making it harder for edge \num{3} to transition upwards.
Conversely, in the \textit{101} state, the opposite is true: edge \num{2} has a high peak and must therefore sit in the low voltage region, while edge \num{3} has a low peak and can sit at high voltage.
On the ramp down, the \textit{010} state looks the same as on the ramp up.

The network's antiferromagnetic behavior is summarized in the transition sequence graph in Fig.~\ref{fig:encoded}(g).
For all but the lowest values of $V_0$, the network sits in either the \textit{010} or \textit{101} states.
To transition between the two, the network must pass through two unstable states, \textit{110} and \textit{100} on the way up, and \textit{001} and \textit{011} on the way down.
The state \textit{111} is not reachable from any point on this graph, highlighting an important deviation from the interacting hysteron model \cite{teunisse2024transition}. 

While the network presented here demonstrates anti-cooperative interactions, the circuitry highlighted in Fig.~\ref{fig:encoded}(a) can easily be modified for cooperative or mixed interactions.
Supplemental Information, Section~\ref{sec:ferro} demonstrates that encoded cooperative interactions can lead to avalanches of the form ``up-up" in SPICE simulation.
Furthermore, the network-dependent gate voltage framework can be extended beyond nearest-neighbor interactions.
The versatility and ease of construction makes this platform ideal for probing systems with memory that extend beyond the hysteron model. 

\section{Discussion}
\label{sec:discussion}

We have presented an experimentally viable electronic element that contains a region of negative differential resistance, giving it the property of bistability. 
Furthermore, our device's IV curve is strongly tunable via two tuning parameters, a gate voltage and linear resistance.
This is one of the few \cite{martinez2024fluidic, brandenbourger2020tunable, geddes2010bistability} constructions of a bistable system in a flow network, and the only one with such strong tunability as to be able to impart precise control over element properties, and therefore over network behavior. 
We have demonstrated how to geometrically construct solutions to these network equations for arbitrary network topologies. 
As a proof of concept of the system's network-level memory capabilities, we showed how to recover any configuration of binary edge states from the history of the global voltage alone.

We then investigated the behavior of our network that extends beyond what can be explained with the hysteron model. 
In a network of otherwise independent bistable resistors, effective interactions arise as a result of the geometry and driving protocol of the network, resulting in avalanches \cite{shohat2025geometric}.
We may also impose explicit interactions in our system, by treating a resistor's IV characteristic as dependent on the state of neighboring resistors. 
We demonstrate an interaction encoding that results in a network with antiferromagnetic behavior.
The versatility of our electronic platform allows for encoding interactions of arbitrary form, opening pathways to examine increasingly exotic hysteretic behaviors in an experimental system.

Emergent memory effects arise on the network scale only when elements are tuned relative to each other, implying that inverse design or learning principles \cite{dillavou2024machine, ben2024multistable} could be applied to this system to exhibit a more expressive space of functionality.
Future expansions of our experimental framework could involve building more complex network topologies beyond series circuits, as well as constructing interacting systems that exhibit sustained oscillations from a single DC drive \cite{ruiz2020topologically}. 
The latter may serve as a generalized model for instability-mediated fluid transport in biological flow networks \cite{geddes2010bistability}.

\begin{acknowledgments}
We thank Joseph Paulsen for helpful discussions and insights related to this work. This work was supported by NSF grant MRSEC/DMR-2309043. M.R.-G. acknowledges support from Ramón y Cajal program (RYC2021-032055-I) funded by MCIN/AEI/10.13039/501100011033 and by European Union NextGenerationEU/PRTR, a Research Grant from HFSP (Ref.-No: RGEC33/2024) with the award DOI (https://doi.org/10.52044/HFSP.RGEC332024.pc.gr.194170) and grant PID2023-147067NB-I00 funded by MCIU/AEI/10.13039/501100011033 and by ERDF/EU. EK acknowledges support from the HFSP Award 977405.
\end{acknowledgments}

\bibliographystyle{unsrtnat}
\bibliography{NDR}

\newpage
\section{Methods}
\subsection{Device Construction}

\begin{figure*}
    \centering
    \includegraphics[width=0.9\linewidth]{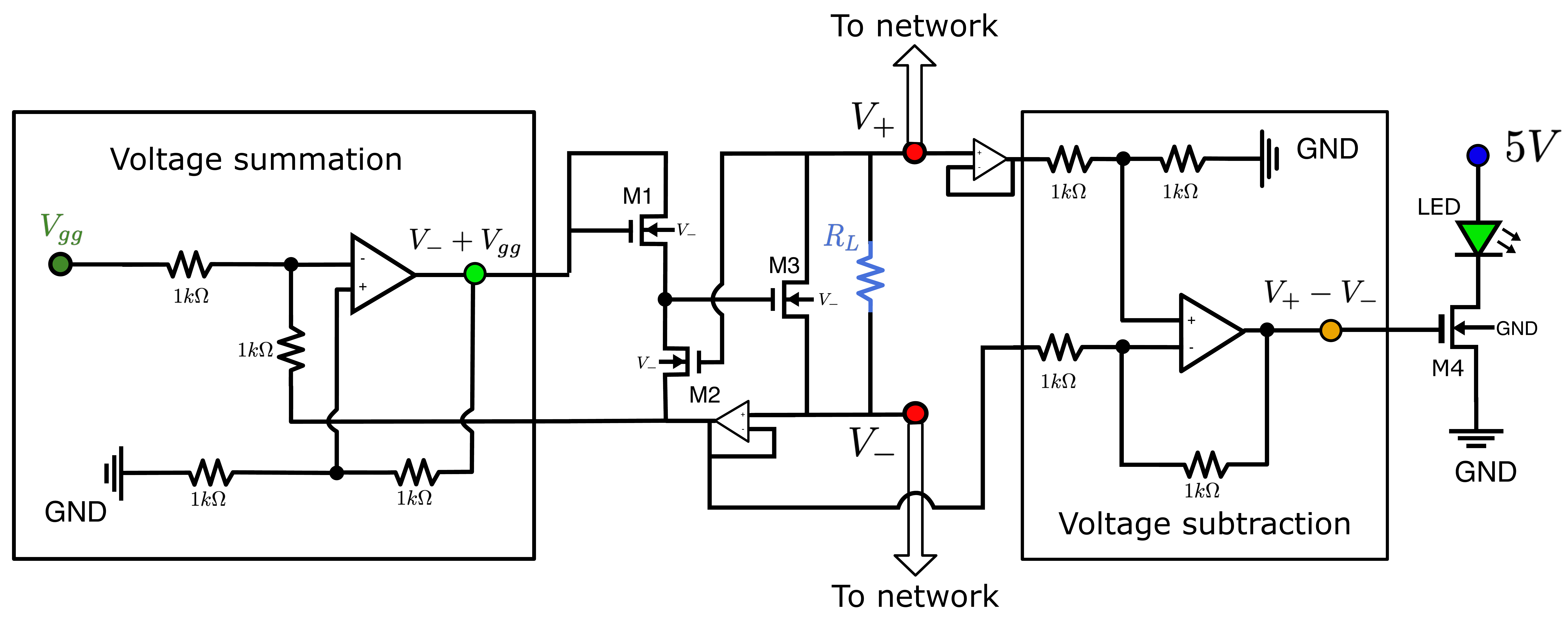}
    \caption{Complete circuit diagram. The primary NDR structure, which includes MOSFETs M1, M2, and M3, receives a gate voltage $V_{gg}$ from a voltage summation circuit (left).
    The two ends of the NDR device, $V_+$ and $V_-$ are buffered and sent to a voltage subtraction circuit (right) to produce $\Delta V = V_+ - V_-$. 
    MOSFET M4 acts as a voltage-controlled current source to power an LED.}
    \label{fig:circuit}
\end{figure*}

The complete circuit diagram of the NDR device used in this work is detailed in Fig.~\ref{fig:circuit}.
One NDR circuit element is defined as an edge in an electronic network, such that it connects nodes $V_+$ and $V_-$ and experiences a current $I$ through those nodes.
N-channel MOSFETs (ALD1106PBL) M1, M2, and M3 act as part of the primary NDR structure \cite{sarkar2016dc} and have their base tied to $V_-$.  
The above construction produces the first positive differential region of the IV curve and the negative differential region, but requires active driving via a DC gate voltage, $V_{gg}$, which is supplied externally.
When $V_{gg} > 3 V_{th}$, with $V_{th}$ as the threshold voltage of each NMOS, the device exhibits the desired non-monotonicity.
By placing this MOSFET configuration in parallel with a linear resistor of resistance $R_L$, we obtain the second positive differential region.
Two operational-amplifiers (OP-AMP) (TLV274IN) are used as voltage followers for $V_+$ and $V_-$, relative to a global ground. 
Un-buffered $V_+$ and $V_-$ are the nodes of the external network.
A third OP-AMP is used for voltage summation to supply the voltage $V_- + V_{gg}$ to the primary circuit as the gate and source of M1. 
The fourth OP-AMP takes the voltage-followed values of $V_+$ and $V_-$ to perform a voltage subtraction, producing $\Delta V = V_+ - V_-$. 
This value is fed as the gate voltage for MOSFET M4 (ALD1106PBL) with its base tied to ground. 
M4 acts as a voltage-controlled current source for the LED, such that the LED's brightness is proportional to $\Delta V$.
Power is supplied to the OP-AMPs of each edge using Korad KD3005D DC power supplies.

\subsection{Experimental supervision}
All experiments were performed using a computer with the assistance of analog data acquisition devices.
The gate voltage, $V_{gg}$, of each edge was applied using an MC USB-3105 Analog Output Device, and linear resistances $R_L$ were modified manually.
The voltage drop across each edge, $\Delta V$, was measured using an MC USB-1608G DAQ Device.
Experiments were controlled with python software using the MCCDAQ package. 
Small networks and individual edges were measured using the Digilent Analog Discovery 2 Oscilloscope and with the Digilent Waveforms software.

\subsection{Binary state generative algorithm}
The network of nine NDR edges that is described in Sec.~\ref{sec:emergent} can be controlled entirely through a ramping protocol of the global source voltage, $V_0$, given that the relative tunings of each edge are known.
The ramping protocol, $V_0(t)$, can be found through an iterative feedback algorithm which runs in a quasi-static manner.
The generative algorithm is described below.

To achieve a desired binary string state, for example, ``010111010", which is depicted in Fig.~\ref{fig:9NDR}(c), and starting with $V_0 = \SI{0}{V}$,

\begin{enumerate}
    \item Take the position of the last ``1" in the series (in this case, position \num{8}). Ramp $V_0$ up until that edge transitions into the on state. 
    \item Remove all ``1"s from the binary string that follow the last ``0". The new string in this example would be ``0101110."
    \item Take the position of the last ``0" in the series (in this case, position \num{7}). Ramp $V_0$ down until that edge transitions into the off state.
    \item Remove all ``0"s from the binary string that follow the last ``1". The new string in this example would be ``010111."
    \item Repeat until the string is empty.
\end{enumerate}

This algorithm does not require exact knowledge of the IV characteristics of each edge.
The only information required is the assumption that the switching ordering of edges is as described in Eq.~\ref{eq:current_order}, and that there is a well-defined threshold voltage for determining when an edge has transitioned from one state to the other.

\section{Data Availability}
The data that support the findings of this study are available via FigShare at \url{https://doi.org/10.6084/m9.figshare.28706885}.

\section{Code Availability}
The code that supports the findings of this study is available on Code Ocean at \url{https://codeocean.com/capsule/5161844}.



\clearpage


\setcounter{figure}{0}
\setcounter{section}{0}
\renewcommand{\figurename}{Fig.}
\renewcommand{\thefigure}{S\arabic{figure}}
\renewcommand{\theequation}{S.\arabic{equation}}
\renewcommand{\thesection}{S\Roman{section}}

\maketitle

\section{Preisach Model}
\label{sec:preisach}
The transition states of non-interacting hysterons are defined according to the Preisach model \cite{mayergoyz1988generalized}, which is generalized for any object with an input-output relationship that contains a multi-branch nonlinearity. 
Transitions between branches occur at extrema of this relationship, and for a network of hysterons all subject to the same global field $H$, one may write down switching field values for each hysteron in both the positive and negative direction, $h_i^+$ and $h_i^-$, such that when $H$ is increasing, the next hysteron to transition to its upper branch will be the one with the minimum positive switching field among those hysterons that are currently in the lower branch, $\text{argmin}(h_i^+)$. 
Conversely, when $H$ is decreasing, the next hysteron to transition to its lower branch will be the one with the maximum negative switching field among those hysterons that are currently in the upper branch, $\text{argmax}(h_i^-)$. 
Because hysterons do not interact with each other in this model, these switching fields are independent of network state and determined entirely by the properties of the hysteron itself.

\begin{figure*}
    \centering
    \includegraphics[width=0.8\linewidth]{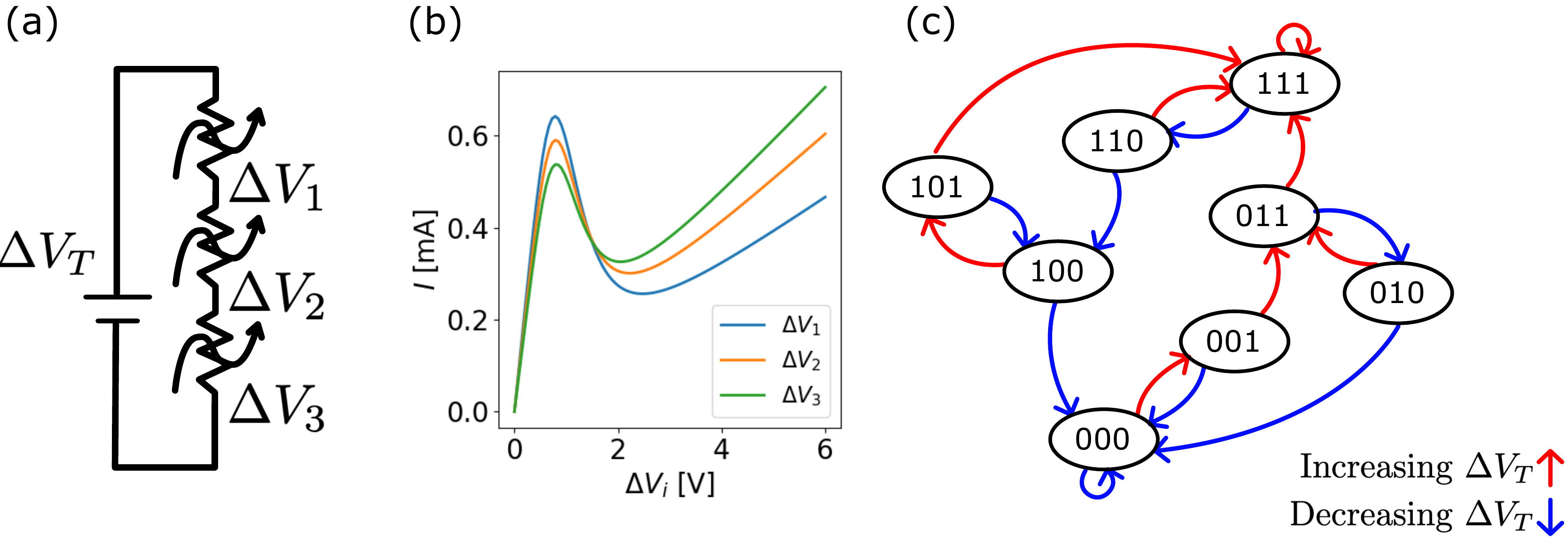}
    \caption{
    Preisach model for transition states of a three-edge network in series. 
    (a) The network diagram shows three nonlinear edges in series, which each have a voltage drop $\Delta V_i$ but share the same current $I$. There is a total voltage drop across the network of $\Delta V_T$.
    (b) IV characteristics of each of the three nonlinear elements. Edge 1's characteristic reaches the lowest peak $I_{max}$ and the highest valley $I_{min}$, and each successive element has a higher peak and lower valley.
    (c) Resulting transition state graph from the peak/valley ordering. Because of the ordering described in (b), all binary string states are reachable from the two saturated states, ``000" and ``111".}
    \label{fig:preisach}
\end{figure*}

In our system of NDR devices in series, such as the one depicted if Fig.~\ref{fig:preisach}(a), the IV curves $I(\Delta V_i)$ are the relevant input-output relationship, and the two positive differential resistance regions are the two branches that are generally accessible because the negative differential resistance region is generally unstable.
The global field here is the current through this network, although we control it indirectly through the source voltage $\Delta V_T$, and the switching fields of each NDR edge are the currents at the peak and valley of its respective IV curve, $I_{max}^i$ and $I_{min}^i$. 
The IV curves depicted in Fig.~\ref{fig:preisach}(b) are such that $I_{max}^1 > I_{max}^2 > I_{max}^3$ and $I_{min}^1 < I_{min}^2 < I_{min}^3$.
In this tuning configuration, the edges will switch up in sequential order and switch down in the same sequential order, giving rise to the transition graph depicted in Fig.~\ref{fig:preisach}(c).
With this transition graph, every binary string state is accessible from the two saturated states, ``000" and ``111".
It should be noted that, while in this demonstration we have placed the devices in geometric order along the series chain to correspond with their switching order, their physical placement in the circuit is irrelevant for switching order.

This ordering of positive and negative switching fields may be extended to any number of devices in series, although drawing the transition graphs becomes cumbersome. 
The demonstration in Fig.~\ref{fig:9NDR} obeys the same ordering, and therefore can access any binary string state by modifying the source voltage alone.

It was pointed out by \cite{shohat2025geometric} that a network of elements such as ours in series under voltage control would not obey the Priesach model explicitly, but would rather exhibit geometric interactions resulting from our indirect control of the true global switching field, $I$.
In a model with interactions, the switching fields of each edge will depend on the network's state, leading to exotic effects like avalanches.
While we do observe avalanches in our system for specific tuning configurations, we find that the non-interacting model is sufficient for the demonstration in Fig.~\ref{fig:9NDR}, and interactions in this regime are negligible.
For more information on avalanches in our system, refer to the Discussion and Supplemental Video~3.

\section{Effective Interaction Terms}
\label{sec:9ndr_interactions}

The demonstration in Sec.~\ref{sec:emergent} depicts nine NDR elements in series that have been tuned to obey a specific state sequence. 
We note that, because the network is in series under voltage control, there arise effective interactions of the same type as described in Sec.~\ref{sec:avalanche}. 
We argue, however, that these interactions are not strong enough to cause a significant change in network behavior, and in particular do not affect the transition sequence. 
This is in alignment with previous findings \cite{shohat2025geometric,liu2024controlled} which state that the strength of interactions scales as $1/N$, with $N$ as the number of bistable elements.

\begin{figure*}
    \centering
    \includegraphics[width=0.7\linewidth]{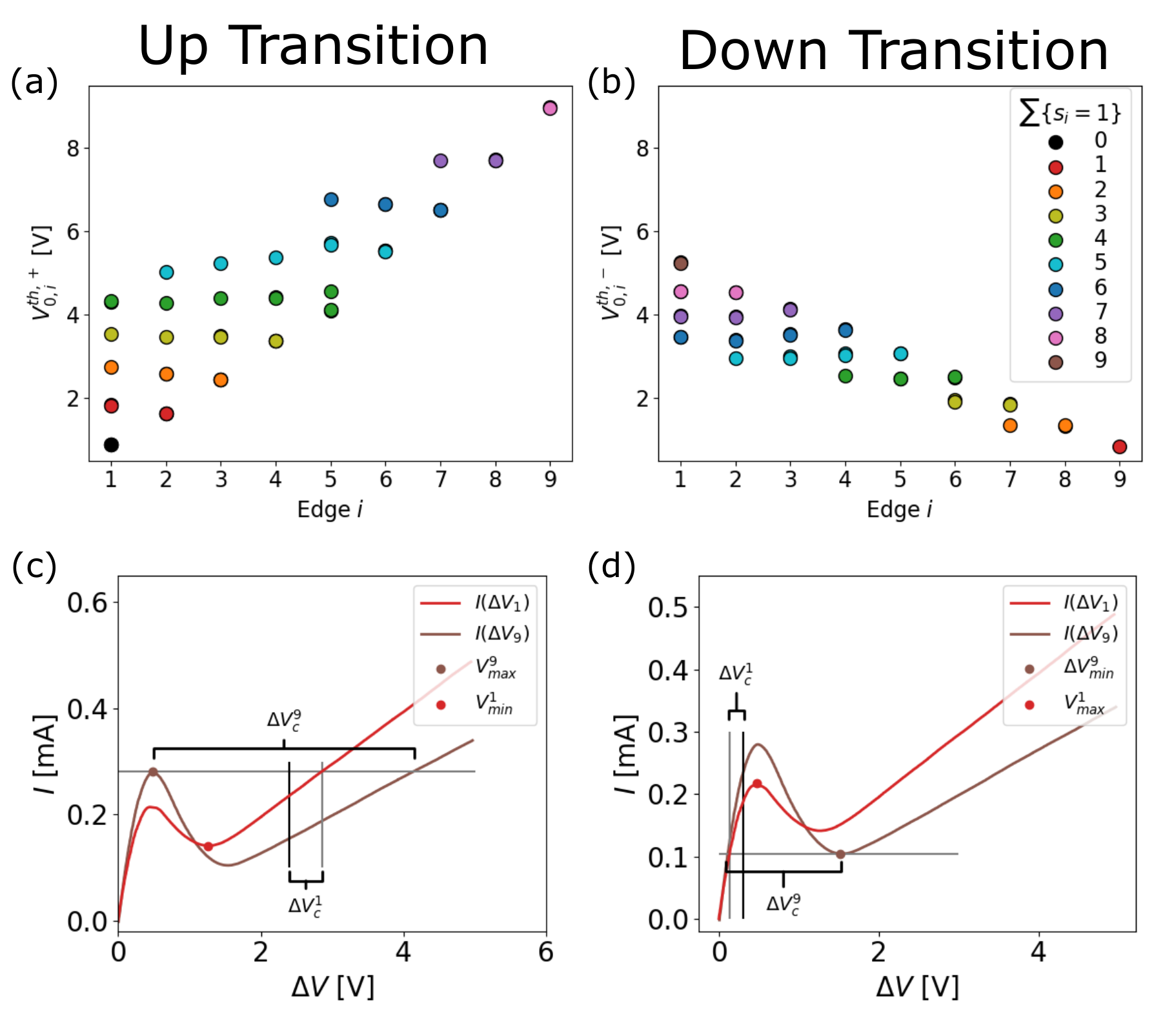}
    \caption{Visualized interactions for the network of nine edges in series depicted in Fig.~\ref{fig:9NDR}. (a-b) Transition thresholds of the global drive, $V_0$, for each edge.
    (a) Upward transition thresholds $V_{0,i}^{th, +}$.
    (b) Downward transition thresholds $V_{0,i}^{th, -}$. Due to the effective interactions, the transition voltage thresholds for each edge are now dependent on the global system state. Data is obtained for \num{100} transitions from \num{64} of the $512$ possible state sequences. The data is binned by the number of edges that are in the ``1" state at the time of the transition.
    (c-d) Geometric construction of the effective interaction on edge \num{1} as a result of a transition in edge \num{9}.
    (c) Edge \num{1} begins in the ``1" state. Edge \num{9} transitions upward, resulting in a voltage decrease in edge \num{1} of $\Delta V_c^1$. The new $\Delta V_1$ does not drop below $\Delta V_{min}^1$, so there is no avalanche.
    (d) Edge \num{1} begins in the ``0" state. Edge \num{9} transitions downward, resulting in a voltage increase in edge \num{1} of $\Delta V_c^1$. The new $\Delta V_1$ does not exceed $\Delta V_{max}^1$, so there is no avalanche.}
    \label{fig:9interactions}
\end{figure*}

If the system were not interacting, the transition thresholds of each edge $i$, $V_{0, i}^{th, \pm}$, would be constant, regardless of the state of the network.
However, in Fig.~\ref{fig:9interactions}, we see that the thresholds can take on multiple values, dependent on the network's binary state.
Canonically \cite{shohat2025geometric}, the thresholds in an interacting network would take the form
\begin{equation}
    V_{0,i}^{th, \pm} = v_{0,i}^{th, \pm} + \sum_j c_{ij} s_j,
\end{equation}
with $v_{0,i}^{th, \pm}$ as the independent switching threshold, $s_j$ as the binary state of all other edges $j$, and $c_{ij}$ as the matrix of interaction coefficients.

The thresholds presented here were extracted from three experimental ramping sequences: ramping $V_0$ between the two saturating states, and the ramping sequences in Figs.~\ref{fig:9NDR}(e) and (f) to produce the ``$+$" and ``$\times$" patterns.
From these data, we obtained \num{100} transitions for \num{64} of the $2^9 = 512$ possible state sequences, $s_i$.
Because of the large number of different $s_i$, we have binned the data in Fig.~\ref{fig:9interactions} by the number of edges that are in the ``1" state at the time of the transition, and colored the data points accordingly. 

Because we are working from incomplete data, we cannot reliably extract the $c_{ij}$ interaction coefficients from these measurements.
However, we may use a simple geometric argument to show that avalanches will not occur in our network as a result of effective interactions.

Suppose a network of $N$ elements is in the state \textit{\{00...1...00\}} such that all elements are in the ``0" state except for one, and let the element in the ``1" state be denoted edge $i$.
An example of such a situation is depicted in Fig.~\ref{fig:9interactions}(d).
If edge $i$ were to transition downward, then its voltage would drop by:
\begin{equation}
    \Delta V_{\text{c}}^i = \Delta V_{\text{min}}^i - \Delta V^i (I_{\text{min}}^i),
\end{equation}
where $\Delta V_{\text{c}}^i$ is the change in voltage drop that edge $i$ experiences as a result of the transition, $I_\text{min}^i$ and $\Delta V_{\text{min}}^i$ are the current and voltage at the valley of its IV curve, and $\Delta V^i (I_{\text{min}}^i)$ is the voltage in the first branch of the IV curve that satisfies the same current as the valley.

If the slopes of all elements' IV curves are relatively similar, then this voltage change will be distributed evenly across all other edges, such that every other edge $j$ would experience an increase in voltage of 
\begin{equation}
    \Delta V_{\text{c}}^j =  \frac{\Delta V_{\text{c}}^i}{N-1}.
\end{equation}
For this transition to result in an avalanche, the change in voltage drop of an edge must be enough to push it over the peak of its IV curve,
\begin{equation}
\label{eq:avalanchecondition}
    \Delta V^j(I_\text{min}^i) + \Delta V_{\text{c}}^j > \Delta V^j_{\text{max}}.
\end{equation}

Note that switching all ``max" with ``min" would give the same condition for an upward transition from the state \textit{\{11...0...11\}}.
If the condition in \eqref{eq:avalanchecondition} is never satisfied for any edge $i$ transitioning either up or down, then no avalanches will occur as a result of the effective interactions.
In the real system, current does not remain constant across a transition, but any changes in current as a result of equilibration would lessen this effect, so our geometric argument serves as a valid prerequisite condition for the presence of avalanches.

In the case of our system of $N=9$ elements, the edge with the most significant voltage drop would be edge \num{9}. 
A down transition results in a $\Delta V_{\text{c}}^9 = \SI{1.42}{V}$, which is distributed equally into each of the eight other edges. 
This transition has the greatest impact on edge \num{1}, and therefore the strongest potential to result in an avalanche. 
The new $\Delta V_1 = \SI{0.31}{V}$ does not exceed the $\Delta V_{max}^1 = \SI{0.47}{V}$.
The same is true of the upward transition: $\Delta V_{\text{c}}^9 = \SI{-3.67}{V}$ results in a new $\Delta V_1 = {2.40}{V}$, which is still greater than the $\Delta V_{min}^1 = \SI{1.26}{V}$.
The cases of the down and up transitions are displayed geometrically in Fig.~\ref{fig:9interactions}(c), (d), respectively.

\section{The Role of Capacitance}
\label{sec:capacitance}

\begin{figure*}
    \centering
    \includegraphics[width=1\linewidth]{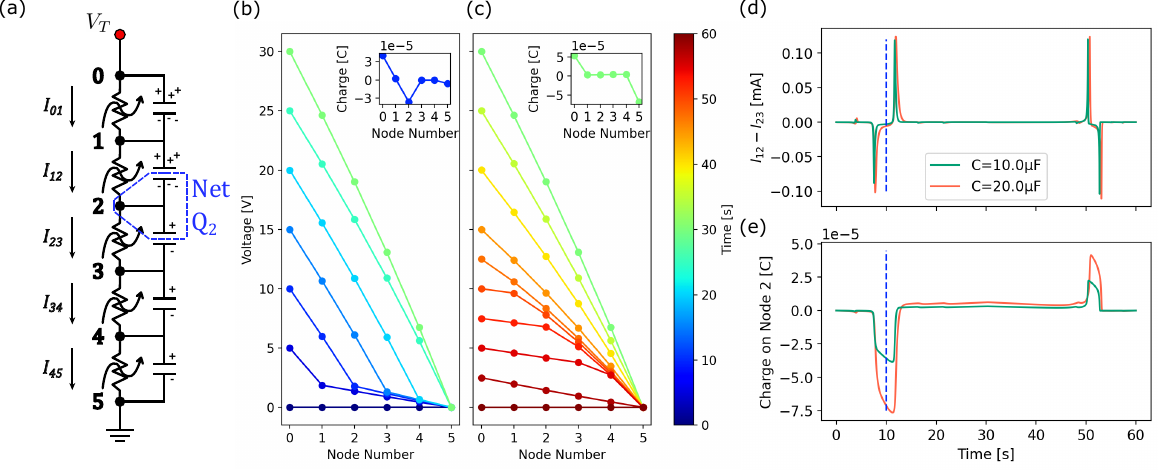}
    \caption{
    The effect of capacitance on network dynamics, as demonstrated with SPICE.
    (a) Circuit schematic. The charge indicated on each capacitor describes the system at time $t=10$~s, when elements 01 and 12 are in the high voltage state and elements 23, 34, and 45 are in the low voltage state. The dotted blue line highlights the resulting net charge on node 2. 
    (b)-(e) SPICE Circuit Simulator data.
    (b) Voltage at each node as the DC source voltage $V_T$ is ramped up from 0 to 30~V. Inset: The net charge on each node at time $t=10$~s.
    (c) Voltage at each node as $V_T$ is ramped down from 30~V to 0. Inset: The net charge on each node when $t=30$; all elements are in the high voltage state, so nodes 1-4 have zero net charge.
    (d) Net current into node 2 as a function of time for $C= \SI{10}{\mu F}$ (green) and $C= \SI{20}{\mu F}$ (orange). Blue dotted line indicated time $t=10$~s.
    (e) Net charge on node 2 as a function of time as calculated by integrating (d) in time.}
    \label{fig:capacitance}
\end{figure*}

Section~\ref{sec:interaction} describes the ideal behavior of a collection of bistable elements that respond to changes in the total voltage applied. 
There is a predicted discontinuous jump in $\Delta V$ as the element switches from the low-voltage state to the high-voltage state (and vice-versa). 
However, a truly discontinuous jump is not physical. 
In fact, a SPICE simulation of the system is unreliable unless a capacitor is added in parallel with each NDR element as shown in Fig.~\ref{fig:capacitance}(a).
The addition of this capacitor satisfies the solver regardless of how small it is, indicating that in the real system, the requirement is satisfied by parasitic capacitance. From the MOSFET data sheet, we estimate this value to be on the order of 1~pF.

To understand how the discontinuous jump and the capacitor are related, consider why we predicted a discontinuous jump: Kirchhoff's laws require that the current through each element in the circuit is always equal. 
What would it mean to relax this constraint? 
If there is a mismatch in the current through two consecutive elements, there must be an accumulation of charge on the wire between them. 
An ideal wire cannot store charge, but any real conductor has some ability to store charge. 
Adding capacitors to the system and considering the effect as their capacitance approaches zero is a good model for the experimental system. 

Consider a system of NDR elements in series where some are in the low-voltage state and some are in the high-voltage state, such as the one depicted in Fig.~\ref{fig:capacitance}(a). 
There are capacitors in parallel across each NDR element, and all of the capacitors are the same size. An electrical node is defined as the point of connection between two or more circuit elements and all the conducting material (\textit{e.g.}, wiring) that is directly connected and therefore held at the same voltage. 
The voltage drop $\Delta V_{ij}$ across the NDR element connecting nodes $i$ and $j$ is calculated as $V_j - V_i$. 
When some NDR elements are in one state and some are in the other, there is a voltage drop mismatch. This can be seen in Fig.~\ref{fig:capacitance}(b) and (c) as the sharp kink that travels across the system over time. 
The charge on a capacitor is proportional to the voltage drop across it, so the capacitors corresponding to the elements in the high-voltage state must be more charged than those corresponding to the elements in the low-voltage state. 
Nodes 1 through 4 each include the negative plate of one capacitor and the positive plate of another, so the net charge on the node between an element in the high-voltage state and an element in the low-voltage state must be nonzero. Consider the governing equation for capacitance:
\begin{equation}
\label{eq:capacticor}
Q_{cap, ij} = C\Delta V_{ij}
\end{equation}
We can use this to arrive at an equation that relates the voltage at each node to the total charge accumulated on it.
\begin{align*}
Q_i &= Q_{cap, i(i+1)} - Q_{cap, (i-1)i} \\
&= C\Delta V_{i(i+1)} - C\Delta V_{(i-1)i} \\
 &= C(2V_i - V_{i-1} - V_{i+1})
\end{align*}
\begin{equation}
\mathbb{Q} = C \hat{L} \mathbb{V},
\end{equation}
where $\mathbb{Q}$ is the vector containing the charges on each node, $\hat{L}$ is the graph laplacian, and $\mathbb{V}$ is the vector containing the voltages at each node. $C\hat{L} = \mathbb{C}$, where $\mathbb{C}$ is the Maxwell capacitance matrix for this system of interacting conductors.

As the total voltage provided to the system is ramped up and down, the charge accumulation travels between nodes. While a node is accumulating charge, the net current into it must be nonzero.
\begin{equation}
\label{eq:ddtcharge}
\frac{dQ_i}{dt} = I_{net, i} = I_{(i-1)i} - I_{i(i+1)}
\end{equation}
Integrating \eqref{eq:ddtcharge} over time provides another way to calculate the charge on each node.
\begin{equation}
Q_i(t) = \int_0^t \left( I_{(i-1)i} - I_{i(i+1)}\right) dt
\end{equation}
All of this hints at a switching timescale that depends directly on the capacitance.
However, the exact form of this timescale is unclear due to the nonlinear nature of the elements. That said, parasitic capacitance is small enough that we can treat our experiments as quasistatic.

For small values of capacitance, the charge appears to move instantaneously from one node to the next. For slightly larger values of capacitance, we can observe a timescale.
Fig.~\ref{fig:capacitance}(d) shows the current mismatch between elements 2 and 3 as a function of time for two different values of capacitance. 
When the capacitance is doubled, the width of the peak increases, indicating that the amount of time it takes for charge to be transferred increases. 
Integrating the net current into a node with respect to time yields the amount of charge stored on that node, as shown in Fig.~\ref{fig:capacitance}(e). Note that the slight deviation from zero between $t=20$~s and $t=40$~s is due to the slight differences in the tuning of the NDR elements.
The total amount of charge stored scales with the size of the capacitors.

\begin{figure*}
    \centering
    \includegraphics[width=0.9\linewidth]{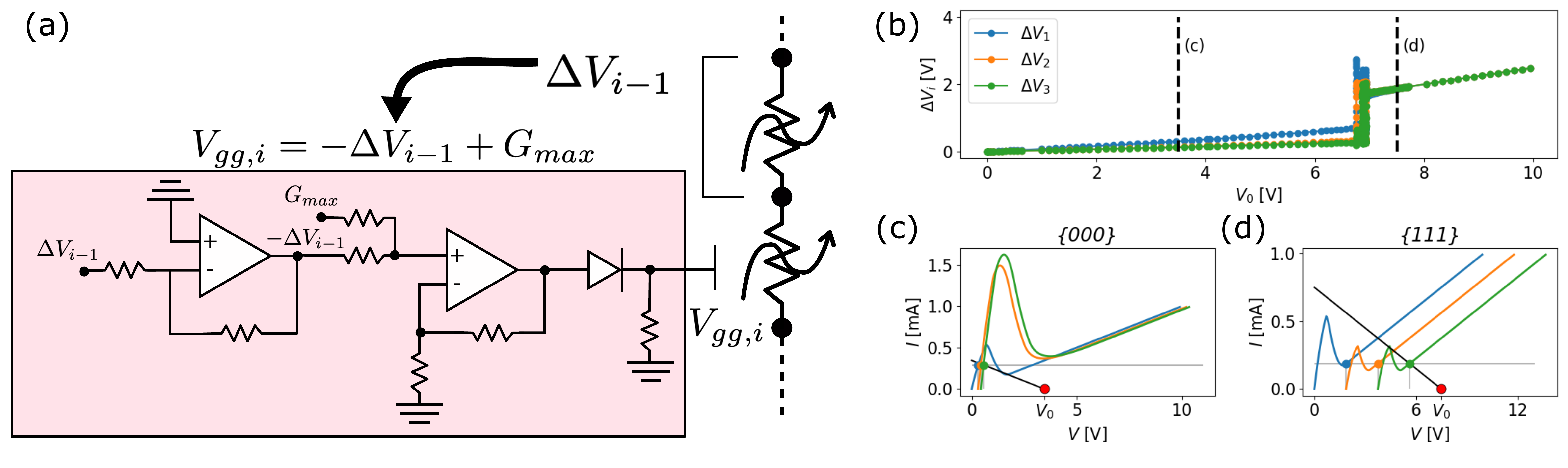}
    \caption{Encoded cooperative interactions result in ferromagnetic behavior in a series circuit.
    (a) Circuitry to implement the varying gate voltage as a function of neighboring voltage drops, $V_{gg,i} = -\Delta V_{i-1} + G_{max}$. The circuitry includes an inverting op-amp, a summing amplifier, and a diode rectifier.
    (b) SPICE simulation data of voltage drops $\Delta V_i$ as a function of global voltage $V_0$ a network with three edges in series with the interactions shown in (a). 
    (c-d) Geometric construction snapshots of the network in the \textit{000} (c) and \textit{111} (d) states.}
    \label{fig:ferro}
\end{figure*}

\section{Encoded Ferromagnetic Interactions}
\label{sec:ferro}

In Sec.~\ref{sec:encoded}, we present a framework for encoding explicit interactions of arbitrary form into networks with bistable elements. 
The demonstration in Fig.~\ref{fig:encoded} displays an encoding for anti-cooperative interactions, leading to a globally antiferromagnetic network behavior. 
Here, we present an alternative encoding with cooperative interactions, leading to globally ferromagnetic behavior.

Fig.~\ref{fig:ferro}(a) shows schematically the circuitry required to encode the cooperative interactions.
Namely, a gate voltage of edge $i$, $V_{gg,i}$, must scale negatively with the voltage drop of its neighbor, $\Delta V_{i-1}$:

\begin{equation}
\label{eq:ferro}
    V_{gg,i} = -\Delta V_{i-1} + G_{max},
\end{equation}

Where $G_{max} = \SI{4}{V}$ is a constant that sets the maximum gate voltage of the edge.
The circuitry to implement this involves an op-amp inverter, a voltage adder, and a diode rectifier.
The rectifier is present to ensure that gate voltages do not drop below \SI{0}{V}.

We now construct a simulated network with this interaction relationship using LTSPICE. 
The network structure is much the same as that of the antiferromagnetic network. 
Three nonlinear resistors are placed in series with one linear resistor, $R_0 = \SI{10}{k\Omega}$, and the network is controlled via a global source voltage, $V_0$.
The first nonlinear resistor is held at a constant gate voltage, $V_{gg,1} = \SI{2.5}{V}$, while the gates of edges \num{2} and \num{3} obey the relationship in Eq.~\eqref{eq:ferro}.
It should be noted that this relationship is implemented ideally in the SPICE simulation, rather than using the circuitry presented in Fig.~\ref{fig:ferro}(a).

Fig.~\ref{fig:ferro}(b) shows the voltage drops of each element, $\Delta V_i$, as $V_0$ is ramped upwards.
Initially, the network is in the \textit{000} state with the IV curves of edges \num{2} and \num{3} having very high peaks, as can be seen in the geometric construction in Fig.~\ref{fig:ferro}(c). 
When $V_0 = \SI{6.76}{V}$, Edge \num{1} undergoes an up transition, leading to cascading transitions for edges \num{2} and \num{3} to transition upwards.
After the avalanche, the network is in the \textit{111} state, owing primarily to the lower peak height for edges \num{2} and \num{3}, as can be seen in Fig.~\ref{fig:ferro}(d).

Cooperative interactions cannot be achieved with effective interactions alone.
Encoded interactions are necessary for avalanches of the form ``up-up," like the one shown here. 
We have therefore demonstrated a novel behavior 

\section{Supplemental Video~1: Geometric solution for the current passing through a small network}
A NDR device is in series with a linear resistor, subject to a total voltage drop $\Delta V_T$.
As $\Delta V_T$ increases, the NDR moves from the low-voltage branch to a region of multistability, but only transitions to the high-voltage branch when that is the singly stable state, at sufficiently high $\Delta V_T$.
Conversely, when $\Delta V_T$ is decreasing, the NDR transitions down to the low-voltage branch once it has left the multistable region.

\section{Supplemental Video~2: Collective memory in a tuned network of nine nonlinear elements}
The total source voltage $V_0$ is measured by a multimeter.
First, $V_0$ is ramped from \SI{0}{V} to \SI{28}{V} then back down to \SI{0}{V}. 
The LEDs of each element turn on in order from \num{1} to \num{9} while $V_0$ is increasing, and turn off in the same order while $V_0$ is decreasing, demonstrating the transition state ordering as dictated in Eq.~\eqref{eq:current_order}.

The network undergoes the sequence in Fig.~\ref{fig:9NDR}(e) 
to achieve the ``$+$'' state, shown in Fig.~\ref{fig:9NDR}(c). 
The source voltage is then reset to \SI{0}{V} and the network then undergoes the sequence in Fig.~\ref{fig:9NDR}(f) to achieve the ``$\times$'' state, shown in Fig.~\ref{fig:9NDR}(d). 

\end{document}